\documentclass[prl,twocolumn,showpacs]{revtex4}
\usepackage{graphicx}
\usepackage{amssymb,amsmath}
\usepackage[colorlinks=true,urlcolor=blue,citecolor=green]{hyperref}

\begin{document}
\title{A versatile simulator for specular reflectivity study of multi-layer thin films}
\author{Sirshendu Gayen}
\email{sirshendu.gayen@saha.ac.in, s.gayen@gmail.com}
\affiliation{{\it Surface Physics Division, Saha Institute of
Nuclear Physics, 1/AF Bidhannagar, Kolkata 700 064, India}}
%
%
\begin{abstract}
A versatile X-ray/neutron reflectivity (specular) simulator using
LabVIEW (National Instruments Corp.) for structural study of a
multi-layer thin film having any combination, including the
repetitions, of nano-scale layers of different materials is
presented here (available to download from the link provided at the
end). Inclusion of absorption of individual layers, inter-layer
roughnesses, background counts, beam width, instrumental resolution
and footprint effect due to finite size of the sample makes the
simulated reflectivity close to practical one. The effect of
multiple reflection is compared with simulated curves following the
exact dynamical theory and approximated kinematical theory. The
applicability of further approximation (Born theory) that the
incident angle does not change significantly from one layer to
another due to refraction is also considered. Brief discussion about
reflection from liquid surface/interface and reflectivity study
using polarized neutron are also included as a part of the review.
Auto-correlation function in connection with the data inversion
technique is discussed with possible artifacts for phase-loss
problem. An experimental specular reflectivity data of multi-layer
erbium stearate Langmuir-Blodgett (LB) film is considered to
estimate the parameters by simulating the reflectivity curve.
\end{abstract}
%
%
%
%

\maketitle
\pagebreak

\section{Introduction}
The reflectivity study is a very powerful scattering technique
\cite{parratt, russell, gibaud, tolan, nielsen, physrep} performed
at grazing angle of incidence to study the structure of surface and
interface of layered materials, thin films when the length scale of
interest is in nm regime. Utilizing the intrinsic magnetic dipole
moment of neutron, the neutron reflectivity study provides magnetic
structure in addition with the structural information. In
reflectivity analysis, the electron density (ED)/scattering length
density (SLD) (in case of X-ray/neutron) of different layers along
the depth is estimated by model-fitting the experimental data.
Text-based language like \emph{FORTRAN} is commonly used to write a
simulation and data analysis code. However, recently a graphical
language LabVIEW ((\textit{Lab}oratory \textit{V}irtual
\textit{I}nstrument \textit{E}ngineering \textit{W}orkbench, from
National Instruments Corp.) has emerged as a powerful programming
tool for instrument control, data acquisition and analysis. It
offers an ingenious graphical interface and code flexibility thereby
significantly reducing the programming time. The user-friendly and
interactive platform of LabVIEW is utilized, here, to simulate a
versatile angle-resolved reflectivity at glancing angles. There is
only a recent report \cite{praman} of LabVIEW-based reflectivity
simulator for energy-resolved reflectivity study where the effects
of absorption and interfacial roughnesses are not included. The
\href{http://henke.lbl.gov/optical_constants/multi2.html}{center for
X-ray optics (CXRO)} provides a similar online simulator, however,
the LabVIEW-based simulator has complete freedom to customize the
programme according to user's choice.

Let us start with brief review on theoretical formalisms of specular
reflectivity with LabVIEW simulated curves. Scattering geometry for
specular scan, polarized neutron reflectivity and reflectivity from
liquid surface are discussed in the next section. Data inversion
technique and the phase problem are discussed then considering the
Auto-correlation function (ACF). Finally, the LabVIEW-based
reflectivity simulator is discussed in detail.

\section{Study of grazing-incidence reflectivity with LabVIEW
simulated curves} The basic quantity that is measured in a
scattering experiment \cite{nielsen, sivia} is the fraction of
incident flux (intensity of X-ray/number of neutrons) that emerges
in various directions is known as the differential cross-section.
Normalizing this quantity by the incident intensity and density of
scatterer, one obtains the scattering rate, $R$ which is termed as
reflectivity (merely, differed by some constant in different
convention and dimensionality) in nano-scale study of materials at
grazing angle. For elastic scattering, $R$ depends only on the
momentum transfer, $\overrightarrow{q}$ which can be varied by
varying the energy (energy-resolved reflectivity) of the incident
beam \textit{i.e.} with a white beam at a fixed grazing incident
angle or by varying the incidence angle (angle-resolved
reflectivity) using monochromatic beam. The angle-resolved
reflectivity which is common in practise for its better resolution
is discussed here. In following discussion We consider, mainly,
X-ray reflectivity however the formalism holds exactly the same way
for neutron reflectivity considering SLD instead of ED.

\textbf{The Born Formalism:} In quantum mechanical treatment, the
incident beam, represented by a plane wave emerges out as spherical
wave when interacted with the scattering potential. Incident plane
is related to the emerging spherical wave through the integral
scattering equation. In first Born approximation of the integral
scattering equation, the scattering amplitude depends on the Fourier
transforation (FT) of the scattering potential. When the scatterer
is composed of homogeneous layers parallel to the x-y plane, the
scattering amplitude, in the first-order approximation, depends only
on the FT of the gradient of ED along the z-axis, implying the
expression for reflectivity \cite{parratt, russell, gibaud, tolan,
nielsen, physrep, sivia} as:
\begin{equation}
R(q_z)=\frac{(4\pi
r_e)^2}{q^4_z}\left|\int_{-\infty}^{+\infty}\frac{d\rho(z)}{dz}e^{iq_zz}dz\right|^2
\label{Born}
\end{equation}
where $\rho(z)$ is the ED at depth z (from top of the sample)
averaged over the x-y plane, $r_e=e^2/{m_ec^2}=2.818\times10^{-5}$
{\AA} is classical radius of electron, or the Thomson scattering
length.

\begin{figure}
\includegraphics[width=8 cm]{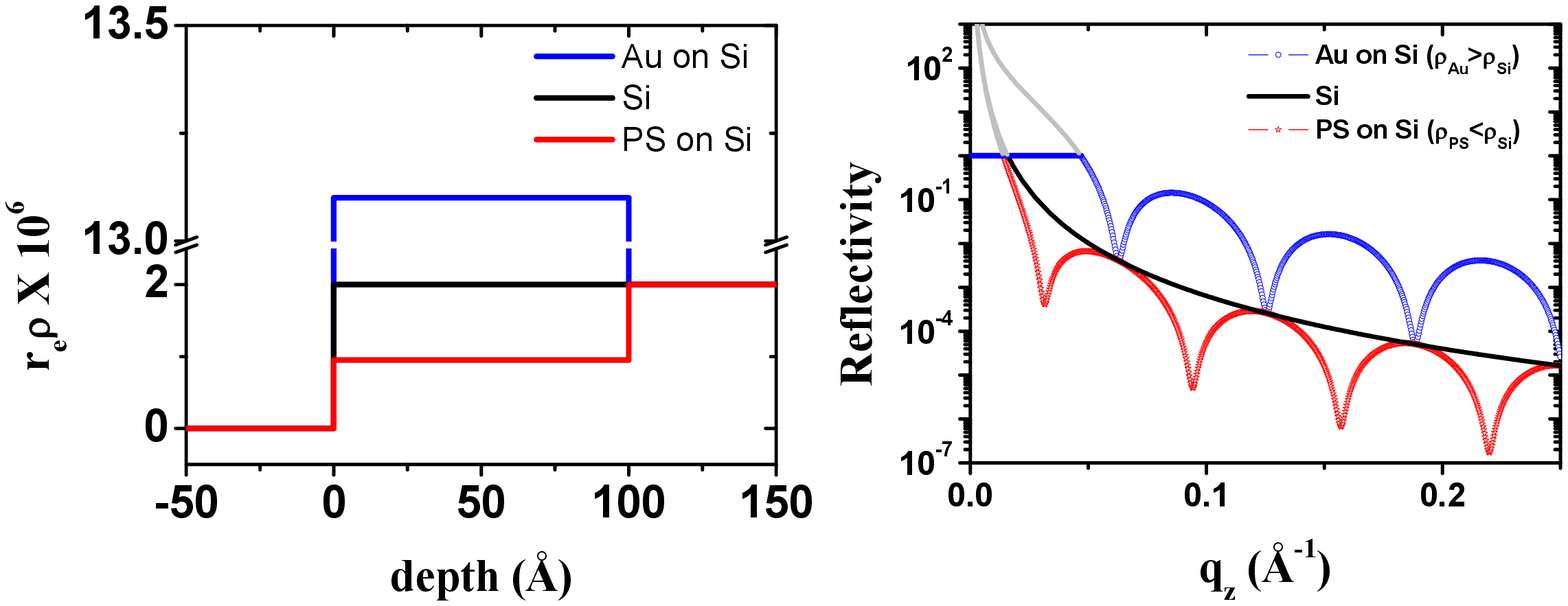}
\caption{\label{Born_refl} The ED profile, $\rho(z)$ and the Born
approximated reflectivity curves for bare Si-substrate (black
curve), a 100 {\AA} thick single layer of Poly-sterene (PS) (red
curve) and Au (blue curve) ($\rho_{Au}>\rho_{Si}>\rho_{PS}$) on
Si-substrate. For small values of $q_z$, the failure of the model is
obvious as  $R$ becomes more than unity (grey curves). The
oscilations (Kiessig fringes) with $\Delta q_z=0.0628$
\text{\AA}$^{-1}$ exactly corresponds to the layer-thickness,
$d=2\pi/{\Delta q_z=}$100 {\AA}. Interesting to note that, the
overall level of the reflectivity curve is higher or lower depending
upon whether the ED of a layer is greater or less than that of the
substrate.}
\end{figure}

In the simplest possible situation where scattering from a bare
substrate is considered, the ED is a step function:

\begin{equation}
\rho(z) =
\begin{cases}
\rho_s &  \text{for } z<0,\\
0 & \text{for } z>0.
\end{cases}
\nonumber
\end{equation}

Obviously,
\begin{equation}
\frac{d\rho}{dz}=-\rho_s\delta(z)
\nonumber
\end{equation}
where $\delta(z)$ is the Dirac's Delta function. Using the integral
property of Delta fn.
\begin{equation}
\int_{-\infty}^{+\infty}\delta(z-0)e^{iq_zz}dz=e^0=1
\nonumber
\end{equation}
Eqn. (\ref{Born}) implies,
\begin{equation}
R(q_z)=\frac{(4\pi r_e\rho_s)^2}{q^4_z}\equiv R_F(q_z)
\label{Fresnel}
\end{equation}

$R_F(q_z)$ is known as the Fresnel reflectivity which does not hold
at small angle as eqn. (\ref{Fresnel}) violates the physical
constraint that $R\leq1$ as $q_z\rightarrow0$. This limitation is a
consequence of neglecting the higher order terms of the scattering
integral equation in first-order Born approximation. Using eqn.
(\ref{Fresnel}), one can rewrite eqn. (\ref{Born}) as
\begin{equation}
R(q_z)=
R_F(q_z)\left|\frac{1}{\rho_s}\int_{-\infty}^{+\infty}\frac{d\rho(z)}{dz}e^{iq_zz}dz\right|^2
\label{Born1}
\end{equation}

When there is a uniform layer of thickness, $d$ on a substrate
(refer to Fig. \ref{Born_refl}), the ED are given by:

\begin{equation}
\rho(z) =
\begin{cases}
\rho_s &  \text{for } z<-d,\\
\rho_1 &   -d<z<0,\\
0 &  \text{for } z>0.
\end{cases}
\nonumber
\end{equation}

whose derivative is a pair of scaled $\delta$-functions:
\begin{equation}
\frac{d\rho}{dz}=(\rho_1-\rho_s)\delta(z+d)-\rho_1\delta(z)
\nonumber
\end{equation}

The integral property of $\delta$-function leads to
\begin{equation}
\int_{-\infty}^{+\infty}\delta(z-0)e^{iq_zz}dz=(\rho_1-\rho_s)e^{idq_z}-\rho_1
\nonumber
\end{equation}

The reflectivity in this case, on simplification is obtained as
\begin{equation}
R(q_z)=\frac{(4\pi
r_e)^2}{q^4_z}\left[\rho^2_1+(\rho_1-\rho_s)^2-2\rho_1(\rho_1-\rho_s)cos(dq_z)\right]
\end{equation}

Superposed on the Fresnel reflectivity ($\propto q_z^{-4}$), it is a
sinusoidal curve with a repeat distance of
\begin{equation}
\Delta q_z=\frac{2\pi}{d}
\nonumber
\end{equation}.

In Fig. \ref{Born_refl}, the black curve is the generated Fresnel
curve following Born approximation for Si
($r_e\rho_{Si}=2.0\times10^{-6}$ \AA$^{-2}$) while the red and blue
curves corresponding to a uniform 100 {\AA} layer of Poly-sterene
(PS) and Au ($\rho_{Au}>\rho_{Si}>\rho_{PS}$) on Si, respectively.
The failure of Born approximation is obvious for small values of
$q_z$ as $R$ blows up beyond 1 (grey region). It is interesting to
note that, the overall level of the reflectivity curve is higher or
lower depending upon whether the ED of a layer is greater or less
than that of the substrate. Moreover, the distinct oscillations
(Kiessig fringes) with $\Delta q_z=0.0628$ \text{\AA}$^{-1}$ exactly
corresponds to the layer-thickness, $d=2\pi/{\Delta q_z=}$100 {\AA}.

\textsl{Generation of Born reflectivity, discrete FT and conversion
of length-scales:}
 The generation of Born reflectivity curve is not
straightforward following eqn.\eqref{Born} which in its analytic
form demands $z$ to be spanned from $-\infty$ to $+\infty$ whereas
in reality the accessible range of $z$ is finite. Hence, one needs
to perform the discrete differentiation and FT for a given range of
$z$ (say, from $z_{min}$ to $z_{max}$ with uniform interval of
$\delta z$) to find $R(q_z)$ over a desired range of $q_z$ (say,
from $q_{min}$ to $q_{max}$ with interval of $\delta q_z$). It may
be mentioned here that what matters in generating the reflectivity
curve is the peaks of $d\rho/dz$ corresponding to each interfaces
(starting from the air-film interface to film-substrate interface)
and the relative separation of the peaks. The last peak in
$d\rho/dz$ corresponding to assumed finite extend (say, 100 $\AA$)
of the substrate does not matter significantly. It is important to
note that one can always perform a coordinate shift in the z-space
and also may redefining the z-range (minimal is the film thickness
as $\rho$ varies in between only) with suitable $\delta z$.
Moreover, the selection of $q_z$ values is free to be selected as
per interest when one performs the FT manually (following eqn.
\eqref{FT}) and $q_{min}$ or $q_{max}$ may have no connection with
$z_{min}$ and $z_{max}$. For a given set of layer-thicknesses ($d$)
and corresponding EDs ($\rho$), it is easy to interpolate the ED
profile, $\rho(z_j)\equiv \rho_j$. Practical surfaces and interfaces
are always rough and the effect of roughness may be included in
$\rho (z)$. When the height variation at any surface or interface is
assumed to be Gaussian, the EDP will be an error-function profile
given by 
\begin{eqnarray}
\rho_j(z)=(\frac{\rho_{j-1}-\rho_j}{2})erfc(\frac{z}{\sqrt{2}\sigma_{j-1,j}})+\rho_j+\nonumber\\
(\frac{\rho_j-\rho_{j+1}}{2})erfc(\frac{d-z}{\sqrt{2}\sigma_{j,j+1}})\\\nonumber
\label{rough}
\end{eqnarray}
where $erfc(x)=\frac{2}{\sqrt{\pi}}\int_x^{\infty}e^{-t^2}dt$. The
discrete differentiation is performed following the \textit{Forward}
method, \textit{i.e.}
\begin{equation}
\frac{d\rho}{dz}|_{z_j}\equiv\rho_j^{\prime}=\frac{1}{\delta
z}(\rho_{j+1}-\rho_j)
\end{equation}
for j =0, 1, . . . , N-1 (considering
$\rho_{N-1}=\rho_N=\rho_{substrate}$) or the \textit{Backward}
method, \textit{i.e.}
\begin{equation}
\rho_j^{\prime}=\frac{1}{\delta z}(\rho_j-\rho_{j-1})
\end{equation}
for j = 0, 1, . . . , N-1 (considering
$\rho_{-1}=\rho_0=\rho_{air}=0$). And the discrete FT is done using
the following method:
\begin{equation}
\mathfrak{r}(q_k)=\sum_{j=0}^{N-1}\rho_je^{-iq_kj}\label{FT}
\end{equation}
where $q_k$ is a given value of $q_z$ and
\begin{equation}
R(q_k)=\frac{(4\pi r_e)^2}{q^4_z}\left|\mathfrak{r}(q_k)\right|^2
\end{equation}
So, for a set of $q_z$ values, one obtains the reflectivity curve.
However, when a standard discrete FT transformer tool or library
function (like a \textit{black-box}) which takes N number (a
convenient choice is a power-of-two \textit{i.e.} $N=2^M$ where $M$
is a positive integer which reduces the computation time if
\textit{fast FT (FFT)} algorithm is used) of $\rho_j^{\prime}$
values as input is used, the usual output is spanned over $\pm\pi =
\pm q_{max}$ with $\delta q=\pi/N$ and one may need to shift the
positive half and the negative half. For standard FT tool with the
exponential in the form of $e^{\pm iq_zz}$ the following relation is
useful for conversion of the length-scales $\delta q_z\delta
z=\pi/N$ (if the exponential has the form of $e^{\pm i2\pi q_zz}$,
then $\delta q_z\delta z=1/N$).

\textbf{The scattering geometry:} To be more precise, eqn.
(\ref{Born}) is the expression for specular reflectivity where the
measurement is done in a $\theta-2\theta$ geometry \textit{i.e.} w.
r. to the beam, sample is rotated by an angle $\theta$ following a
rotation of detector by $2\theta$. One may define the axes by
coinciding a particular axis with the beam direction, in particular
when the beam-direction is fixed in space (for synchrotron
beam-lines), however, as we are interested in structure of the
sample it is convenient to define the axes fixed with the sample.

\begin{figure}
\begin{center}
\includegraphics[width=8 cm]{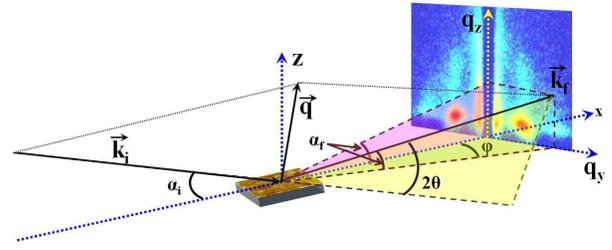}
\caption{\label{scatgeo} The scattering geometry \textit{w. r. to}
the reference axes fixed with the sample. A beam (X-ray/neutron)
incidents at an angle $\alpha_i$ and gets scattered
(reflected/diffracted) towards ($\alpha_f,\phi$). $2\theta$
represents the deviation of the beam from its incident direction.}
\end{center}
\end{figure}

Fig. \ref{scatgeo} shows a general scattering geometry when
reference frame is chosen fixed with the sample. In this geometry,
the incident wave vector is given by,\\
\begin{equation}
\overrightarrow{k_i}\doteq\frac{2\pi}{\lambda}\left(
\begin{array}{ccc}
cos\alpha_i \\
0 \\
-sin\alpha_i\end{array} \right)
\nonumber
\end{equation}
and the scattered wave vector for an angle $\phi$ out-of the plane
of reflection is given by,
\begin{equation}
\overrightarrow{k_f}\doteq\frac{2\pi}{\lambda}\left(
\begin{array}{ccc}
cos\alpha_f cos\phi \\
cos\alpha_f sin\phi \\
sin\alpha_f \end{array} \right)
\nonumber
\end{equation}
where $\alpha_i, \alpha_f$ are the incident and scattered angle w.
r. to the sample. For elastic scattering,
$|\overrightarrow{k_i}|=|\overrightarrow{k_f}|=k=\frac{2\pi}{\lambda}$.
The transfer wave vector,\\
\begin{equation}
\overrightarrow{q}=\overrightarrow{k_f}-\overrightarrow{k_i}\doteq\frac{2\pi}{\lambda}\left(
\begin{array}{ccc}
cos\alpha_f cos\phi- cos\alpha_i \\
cos\alpha_f sin\phi \\
sin\alpha_f + sin\alpha_i \end{array} \right)
\label{q1}
\end{equation}

Using the relation
\begin{equation}
\widehat{k_f}\cdot\widehat{k_i}=cos2\theta
=cos\alpha_icos\alpha_fcos\phi - sin\alpha_isin\alpha_f \nonumber
\end{equation}
one can obtain the following form for magnitude of the momentum
transfer,
\begin{eqnarray}
\nonumber |\overrightarrow{q}|=q
&=&\frac{2\pi}{\lambda}\sqrt{2[1-(cos\alpha_icos\alpha_fcos\phi-
sin\alpha_isin\alpha_f)]}\\
\nonumber &=&\frac{2\pi}{\lambda}\sqrt{2(1-cos2\theta)}\\
&=&\frac{4\pi}{\lambda}sin\theta \label{n}
\end{eqnarray}
Here, one may note that the last expression is equivalent to the
Bragg's condition for $d=\frac{2\pi}{q}=\frac{\lambda}{2sin\theta}$
where $2\theta$ is the deviation of the scattered beam from its
initial direction. One important point is that the expression for
$q_z$, derived from simple vector geometry becomes equivalent,
apparently, with the Bragg's condition which is essentially involved
with interference! Actually, when one considers $d=\frac{2\pi}{q}$,
the essence of interference effect automatically comes in as the
$d-space$ is connected to the $q-space$ by Fourier transformation
and vice-versa. Another point, obvious from the Bragg's condition,
$2dsin\theta=n\lambda$, is that when the small length-scale (atomic
spacing of few {\AA}) is probed, the Bragg peaks should appear at
larger angles ($\theta\sim$ few tens of degrees) whereas for larger
length-scales (few tens or hundreds of nm) the peaks should appear
at smaller angles ($\theta\sim$ few degrees only) --- the former one
is the case of diffraction and the later for reflectivity.

Sometimes, it is convenient to express $\overrightarrow{q}$ in terms
of the parallel (to sample surface) component,
$q_{||}=\sqrt{q_x^2+q_y^2}$ and the perpendicular, $q_{\bot}=q_z$.
The specular condition, $\alpha_i=\alpha_f=\theta$ with $\phi=0$,
implies $q=q_{\bot}=q_z=2k_z=4\pi sin\theta/\lambda$. So specular
reflectivity can be done either by varying $\lambda$ for fixed
$\theta$ (energy resolved) or by varying $\theta$ for fixed
$\lambda$ (angle resolved).\\

\textsl{Footprint correction:} One important correction needs to be
included, particularly for specular reflectivity at small angles,
when the footprint of the beam exceeds the sample size, L. If the
beam width is W, then the footprint, F for an incidence angle of
$\alpha_i$ is given by F=W/sin$\alpha_i$, hence, for F$>$L the
experimental data need to be corrected by multiplying a factor of
F/L.\\

\textsl{Instrumental resolution:} Another crucial point that one has
to consider is the instrument resolution function. The effect of
instrumental resolution can be considered as convolution by the
resolution function which in most cases is approximated by a
Gaussian whose standard deviation is usually determined by the
full-width at half-maxima (FWHM) of the direct beam in a detector
scan. We consider a Gaussian resolution function with suitably
defined window [$\equiv$ (2 $\times$ N$_{resolution}$ + 1) data
points] for post-processing (weighted sum ) of the simulated data to
include the instrumental resolution effect.

\textsl{Specular reflectivity from liquid surface/interface}:
\begin{figure}
\begin{center}
\includegraphics[width=8 cm]{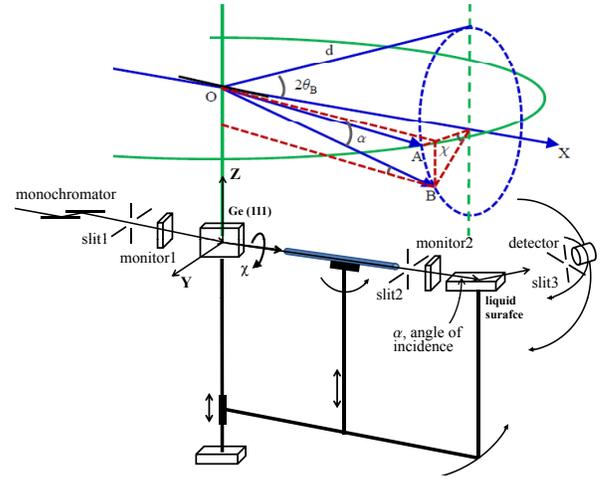}
\caption{\label{liqrefl}Schematic of the experimental set up and the
beam geometry for reflectivity study from liquid surface.
$\protect\overrightarrow{OX}$ is the direct beam while OAX defines
the horizontal plane. The rotation of the deflector crystal (black
line at O) about the direct beam defines a cone of semi-opening
angle $2\theta_B$ at O. The rotation of deflector crystal, $\chi$ is
related to the angle of incidence, $\alpha$ through the relation
$sin\alpha=sin(2\theta_B)sin\chi$. To bring the point of interest
back to its original position a vertically shift of $dtan\alpha$
followed by a horizontally rotation by an angle of
$sin^{-1}[sin(2\theta_B)(1-cos\chi)]$ is required.}
\end{center}
\end{figure}
While discussing about the scattering geometry and experimental set
up for reflectivity study of solid samples, we may have a glimpse on
the reflectivity study from liquid surface or liquid-liquid
interface \cite{liquid1} which is special because the  liquid
surface necessarily be horizontal. Moreover, the reflected intensity
from liquid surface/interface drops drastically with increase in
angle hence a synchrotron beam is preferred. Unfortunately,
synchrotron beams are usually fixed in direction and need to incline
by Bragg-reflection using suitable single-crystal. Usually the
deflector crystal is mounted at O,the center of the goniometer stage
where the solid samples are mounted and the liquids are placed on an
additional stage which can rotate about the vertical axis through O.
The rotation of the deflector crystal about the direct beam
($\overrightarrow{OX}$, refer to Fig. \ref{liqrefl}) obeying the
Bragg condition (for example, the Bragg angle of Ge($111$) at 18 keV
$=\theta_B=6.008^\circ$) causes the Bragg-reflected beam to describe
a cone of opening angle $4\theta_B$
($\sqrt{z^2+y^2}=xtan(2\theta_B)$). As a result, the change in angle
of incidence causes the point of incidence to shift hence for each
angle the point of incidence (the liquid stage as a whole) has to be
brought back to its original position. Now, the locus of the beam on
a virtual cylinder ($\sqrt{x^2+y^2}=d$: defined by the rotation of
the liquid stage about the vertical axis through O) of radius d is
given by $x^2[2-sec^2(2\theta_B)]+2y^2+z^2=d^2$. If a rotation of
the deflector crystal by an amount $\chi$ brings the horizontal beam
$\overrightarrow{OA}$ to $\overrightarrow{OB}$ which makes an angle
$\alpha$ with horizontal (x-y) plane then considering the vertical
shift on a virtual vertical plane (not on the cylindrical surface),
one can write, $dsin\alpha=dsin(2\theta_B)sin\chi$ which relates the
angle of incidence, $\alpha$ to the rotation of the crystal through
the relation $sin\alpha=sin(2\theta_B)sin\chi$. So, for a rotation
of the deflector crystal by an amount of $\chi$, the point of
interest should be vertically shifted by an amount of $dtan\alpha$
(for a line joining the origin and a point $\textbf{r}(x,y,z)$ that
makes an angle $\alpha$ with x-y plane, it is obvious that
$cos\alpha=z/r$ hence $tan\alpha=z/d$) followed by a horizontal
rotation of $sin^{-1}[sin(2\theta_B)(1-cos\chi)]$. To avoid the
shifting of the incident spot while changing the incident angle
two-crystal assembly may also be used where the fixed beam passes
through one focus of an ellipsoid of revolution and the point of
incidence on the liquid surface fixed at another focus while the
driving crystals move in a coupled fashion over two circular
cross-sections of the ellipsoid of revolution. However, using two
crystals involves more loss of beam intensity and more complicity in
alignment.

\textbf{The dynamical theory and Parratt formalism:}
\begin{figure}
\begin{center}
\includegraphics[width=8 cm]{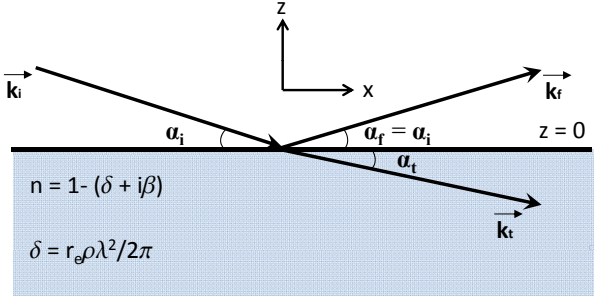}
\caption{\label{fresnel1} Schematic of reflection and refraction of
a wave incident on a smooth surface at an angle $\alpha_i$.
Reflected wave emerges at an angle, $\alpha_f=\alpha_i$ and the
refracted wave transmitted through the medium at an angle
$\alpha_t$. $\protect\overrightarrow{k_i}$,
$\protect\overrightarrow{k_f}$, $\protect\overrightarrow{k_t}$ are
the incident, reflected and transmitted wave vectors, respectively.}
\end{center}
\end{figure}

\begin{figure}
\begin{center}
\includegraphics[width=8 cm]{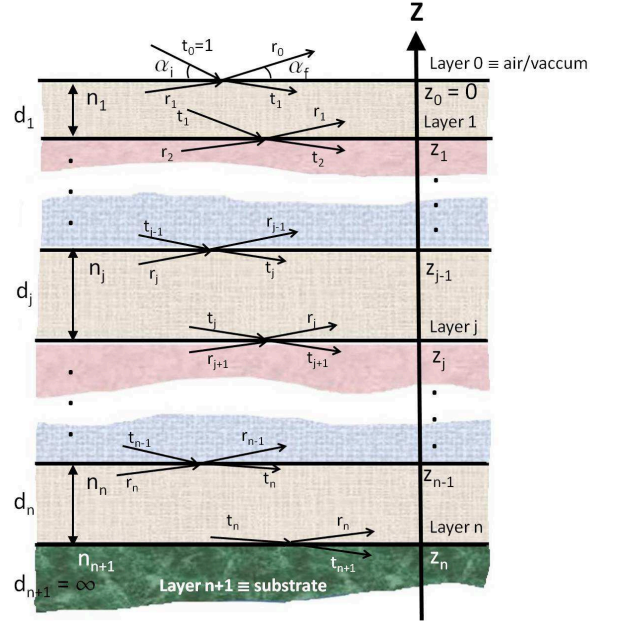}
\caption{\label{multirefl}Schematic of multiple reflection from a
multilayer film having $n$-layers on a substrate with $n+1$
interfaces. For incident wave amplitude is normalized to unity,
$t_0=1$. No reflection from the substrate sets, $r_{n+1}=0$.}
\end{center}
\end{figure}

In classical treatment of scattering, the continuity of the electric
and magnetic field vectors of the propagating electro-magnetic wave
at an interface provides the relation between the reflection ($r$)
and transmission ($t$) coefficients and hence imply the
reflectivity, $R=r^*r$. For a smooth air-medium interface (refer to
Fig. \ref{fresnel1}) with an incidence angle of $\alpha_i=\alpha_f$
(specular), the expression for $r$ and $t$, known as Fresnel formula
is given by
\begin{eqnarray}
r &=&\frac{k_{i,z}-k_{t,z}}{k_{i,z}+k_{t,z}},\\
t &=&\frac{2k_{i,z}}{k_{i,z}+k_{t,z}}
\end{eqnarray}
where
\begin{equation}
k_{i,z}=ksin\alpha_i \nonumber
\end{equation}
and (from Snell-Descartes' law)
\begin{equation}
k_{t,z} = nksin\alpha_t =k\sqrt{n^2-cos^2\alpha_i}\label{t}
\end{equation}
The refracted wave transmits at an angle $\alpha_t$ through the
medium having refractive index,
\begin{equation}
n=1-(\delta+i\beta) \label{n}
\end{equation}
where ED, $\rho$ is redefined as
\begin{equation}
\delta=r_e\rho\lambda^2/2\pi \nonumber
\end{equation}
and $\beta$ is the absorption co-efficient. It is more convenient to
use $\rho$ (electron per ${\AA}^3$) rather than the dimensionless
number $\delta$ because $\rho$ is a $\lambda$-independent specific
number for a material. The dynamical calculation, for a thin film
consisting of a single layer on a substrate, using matrix transfer
method \cite{gibaud, tolan} for each layer which essentially
connects the fields between two consecutive layers, yields the
following relation
\begin{eqnarray}
r &=&
\frac{r_{0,1}+r_{1,2}e^{-2ik_{z,1}}}{1+r_{0,1}r_{1,2}e^{-2ik_{z,1}}}\\
t &=&
\frac{t_{0,1}t_{1,2}e^{-ik_{z,1}}}{1+r_{0,1}r_{1,2}e^{-2ik_{z,1}}}
\end{eqnarray}
where $r_{j,j+1}$ is the coefficient of reflection at the interface
between $j$-th and $(j+1)$-th layer. It may be noted here that the
additive term to unity in the denominator corresponds to multiple
reflection in the film. Extension to a multi-layer film having $n$
homogeneous layers on a substrate, one obtains the recursion
relation for the ratio of reflection and transmission coefficients
of the j-th layer, as follows
\begin{equation}
X_j=\frac{r_j}{t_j}=e^{-iq_{z,j}z_{j}}\frac{r_{j,j+1}+X_{j+1}e^{(iq_{z,j+1}z_{j})}}{1+r_{j,j+1}X_{j+1}e^{(iq_{z,j+1}z_{j})}}
\label{transmatrix}
\end{equation}

where $q_{z,j}=2k_{z,j}$ and $z_j=\sum_{m=0}^j d_m\equiv$ depth
including the j-th layer thickness from top. In our convention
(refer to Fig. \ref{multirefl}), $j=0$ for air/vacuum and $j=n+1$
for the substrate. The expression for $r_{j,j+1}$ is given by
\begin{equation}
r_{j,j+1}=\frac{q_{z,j}-q_{z,j+1}}{q_{z,j}+q_{z,j+1}}
\label{reflectance}
\end{equation}
with
\begin{eqnarray}
q_{z,j}&=&\sqrt{{q_z}^2-\frac{32\pi^2}{\lambda^2}(\delta_{j}+i\beta_{j})}\\
&=& \sqrt{{q_z}^2-16\pi
r_e\rho_j-\frac{i32\pi^2\beta_j}{\lambda^2}}\label{qzj}
\end{eqnarray}
No reflection from the substrate (assumed to be sufficiently thick)
sets
\begin{equation}
X_{n+1}=0
\nonumber
\end{equation}
as the start of the recursion. After $n+1$ iterations, the
expression for specular reflectivity is obtained as
\begin{equation}
R=|X_0|^2 \nonumber
\end{equation}\\

\textsl{Critical Angle:} It may be noted here that, X-ray,
propagating through a medium of higher refractive index, suffers
total external reflections when incident on a surface of a material
having lower refractive index. The very concept of total external
angle defines an angle below which the X-rays are totally reflected
back and termed as Critical angle. Setting $\alpha_t=0$ in eqn.
\eqref{t}, one obtains
$n^2(\approx1-2\delta)=cos^2\alpha_c(\approx1-\alpha^2_c)$
\textit{i.e.} the critical angle, $\alpha_c$ is related to the ED of
the material as
\begin{equation}
\alpha_c\approx\sqrt{2\delta}=\lambda\sqrt{r_e\rho/\pi}=sin^{-1}(\lambda
q_c/4\pi)
\end{equation}
Typical orders of magnitude are: $\delta\simeq 10^{-5}$ and
$\beta\simeq 10^{-6}$ so that $\alpha_c \simeq 0.1^{\circ}$ to
$0.5^{\circ}$. For Si, $\delta_{Si}=7.6\times10^{-6}$ and
corresponding $\alpha_c=0.223^{\circ}$ with $q_c=0.0316$
{\AA}$^{-1}$ for $\lambda = 1.54$ {\AA}. For a multilayer film, the
overall critical angle is determined by the layer having the highest
value of $\delta$. In Fig. \ref{Born_refl}, the critical angle for
blue curve (Au on Si) is defined by $\delta_{Au}$ as
$\delta_{Au}=4.96\times10^{-5}>\delta_{Si}$ whereas for red curve
(PS on Si) the critical angle is defined by $\delta_{Si}$ as
$\delta_{Si}>\delta_{PS}=3.5\times10^{-6}$. For incident angle less
than the critical angle \textit{i.e.} $\alpha_i<\alpha_c$, the
penetration depth is only few nano-meters, however, it increases
sharply to several micro-meters as $\alpha_i$ exceeds $\alpha_c$ and
X-ray immediately sees the whole layer. Actually, there is a
so-called \textit{evanescent wave} within the refracting medium,
propagating parallel to the interface and exponentially decaying
perpendicular to it. In grazing incidence diffraction (GID) study
where only the first few atomic layers are of main interest
$\alpha_i$ is kept just below the $\alpha_c$
\textit{i.e.} ($\alpha_i\lesssim \alpha_c$).\\

The transfer matrix formalism is equivalent to the recursive
approach of Parratt's \cite{parratt, gibaud, tolan} formalism and
both (known as Dynamical Theory) are fairly exact since it
incorporates the issue of multiple scattering. The additive term to
unity in the denominator of eqn. \eqref{transmatrix}, corresponding
to multiplicative reflection \cite{gibaud, tolan}, contributes
significantly for small incident angles and when ignored the
reflectivity expression can be simplified to,
\begin{eqnarray}
\nonumber R(q_z)&=&\left|\sum_{j=0}^nr_{j,j+1}e^{i\sum_{m=0}^j
q_{z,m}d_m}\right|^2\\
&=&\left|\sum_{j=0}^n\frac{q_{z,j}-q_{z,j+1}}{q_{z,j}+q_{z,j+1}}e^{i\sum_{m=0}^j
q_{z,m} d_m}\right|^2 \label{Parratt1}
\end{eqnarray}
where $d_j$ is the thickness of $j$-th layer.

In further assumption that the incident angle does not change
significantly from layer to layer i.e. $q_z$ is same for all layers,
with an additional approximation (so that first two terms of the
binomial expansion of the right hand side of the eqn. \eqref{qzj}
would be sufficient to consider),
\begin{equation}
q_z>q_c=\frac{4\pi}{\lambda}sin\theta_c=\sqrt{16\pi r_e\rho}
\label{3rdbornapprox}
\end{equation}
one can write eqn. \eqref{Parratt1} in a simpler form:
\begin{eqnarray}
R(q_z)&=&\frac{64\pi^4}{\lambda^4q^4_z}\left|\sum_{j=0}^n({\delta_{j+1}-\delta_j})
e^{iq_zz_j}\right|^2\\
&=& \frac{(4\pi
r_e)^2}{q_z^4}\left|\sum_{j=0}^n({\rho_{j+1}-\rho_j})
e^{iq_zz_j}\right|^2 \label{Born2}
\end{eqnarray}
where $\sum_{m=0}^j d_m=z_j\equiv$ depth including the j-th layer
thickness from top. Eqn. (\ref{Born2}) is exactly the same with eqn.
(\ref{Born}) in the continuous limit. This greatly simplified
treatment, known as the Kinematical Theory or Born approximation
reduces the Parratt recursive formula to a simple relation.

In Born approximation, the effect of multiple reflection from
different layers of a multi-layer film is ignored and also the
effect of refraction from one layer to another is not considered by
assuming the incident angles to be same for all interfaces. Hence,
for a bare substrate where there is no multiple reflection and only
one angle of incidence, both the Parratt curve and the Born curve
are expected to be identical and same with the Fresnel curve.
However, the third approximation (\ref{3rdbornapprox}) makes the
difference and fails to restrict the Born curve from blowing up at
small angles.

Formulation of reflectivity expression for scattering of
electromagnetic wave from rough surfaces is difficult as the
solution of the relevant wave equations turns out to be complicated
involving matching of the boundary conditions over random rough
surfaces and several simplifying assumptions have to be invoked for
their solution. An exponential damping factor may be included when
the roughness ($\sigma$) has an error-function profile, as
introduced by N\'{e}vot and Croce \cite{gibaud, tolan} to modify
eqn. (\ref{reflectance}), eqn. (\ref{Parratt1}) and eqn.
(\ref{Born2}) in the following form:

\begin{equation}
r_{j,j+1}=\frac{q_{z,j}-q_{z,j+1}}{q_{z,j}+q_{z,j+1}}e^{-\frac{q_{z,j}q_{z,j+1}\sigma^2_{j,j+1}}{2}}
\label{reflectance2}
\end{equation}
\begin{equation}
R(q_z)=\left|\sum_{j=0}^n\frac{q_{z,j}-q_{z,j+1}}{q_{z,j}+q_{z,j+1}}e^{i\sum_{m=0}^j
q_{z,m} d_m}e^{-\frac{q_{z,j}q_{z,j+1}\sigma^2_{j,j+1}}{2}}\right|^2
\label{Parratt2}
\end{equation}
\begin{eqnarray}
R(q_z)&=&\frac{64\pi^4}{\lambda^4q^4_z}\left|\sum_{j=0}^n({\delta_{j+1}-\delta_j})
e^{iq_zz_j}e^{-\frac{q_z^2\sigma^2_{j, j+1}}{2}}\right|^2\\
&=& \frac{(4\pi
r_e)^2}{q_z^4}\left|\sum_{j=0}^n({\rho_{j+1}-\rho_j})
e^{iq_zz_j}e^{-\frac{q_z^2\sigma^2_{j, j+1}}{2}}\right|^2
\label{Born3}
\end{eqnarray}
It may be noted here that the roughness term as introduce here
carries more weightage for larger $q_z$ hence damps the reflectivity
curve slightly more in comparison with roughness introduced
following eqn. \eqref{rough}.

\begin{figure}
\begin{center}
\includegraphics[width=7 cm]{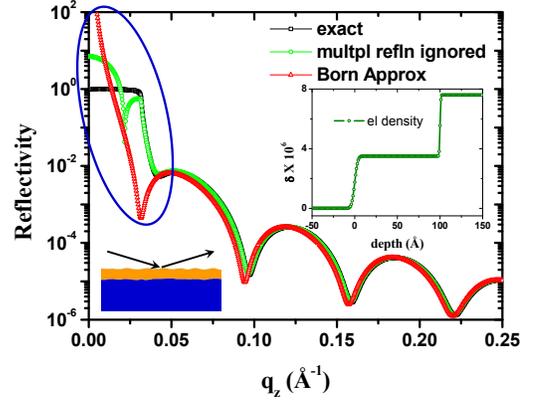}
\caption{\label{compare}Reflectivity curves for a 100 {\AA} thick
single layer of PS on Si substrate following different formalism.
Electron density, convoluted with the roughness is in the inset. The
region highlighted by the blue circle shows distinct difference
between the curves. The back one is for exact dynamical theory. The
effect of multiple reflection is ignored for green curve. Further
approximation (Born) that the incident angle to be same at all
interfaces and $q_z>q_c$ is assumed for the red curve.}
\end{center}
\end{figure}

Fig. \ref{compare} compares the reflectivity curves for a 100 {\AA}
thick single layer of PS on Si substrate following eqn.
(\ref{transmatrix}) using (\ref{reflectance2}), (\ref{Parratt2}) and
(\ref{Born3}). The inset figure shows the electron density profile
with the gaussian roughness (refer to eqn. \eqref{rough}) of the
interfaces. The region highlighted by the blue circle, particularly
around $q_c^{Si}=0.0316$ {\AA}$^{-1}$, shows distinct difference
between the curves. The back one is for exact dynamical theory. The
effect of multiple reflection causes the reflectivity to grow more
than 1 as observed for the green curve. For the Born approximated
(blue) curve, the deviation is even serious as it assumes further
that the
incident angle to be the same at all boundaries.\\
\textbf{Polarized neutron reflectivity (PNR):} In addition to the
structural information, reflectivity study using the spin polarized
neutron (PNR) provides the detail of in-plane magnetic ordering in a
magnetic thin film. Neutrons arriving at the sample surface are spin
polarized either parallel $(+)$ or antiparallel $(-)$ to the
quantization axis defined by the applied magnetic field, H (along
the y-direction, refer to Fig. \ref{scatgeo}). In this case, the
refractive index includes the nuclear and magnetic scattering
contributions and expressed as
\begin{eqnarray}
n^{\pm} (z)=1-(b\mp c\mu)
\label{pnr}
\end{eqnarray}
where $b$ is the complex nuclear SLD and $c=\lambda^2
m_{n}\mathbf{\mu}_{n}/\hbar^{2}$  and $\mathbf{\mu}
(\mu_x,\mu_y,\mu_z)$ is the magnetization. Eqn. \eqref{pnr} is same,
except the additional magnetic term, with eqn. \eqref{n} as $b\equiv
\delta+i\beta$. Depending upon the polarization of incident and
reflected neutron beam, the measured reflectivities are designated
as $R^{++}$, $R^{--}$, $R^{+-}$ and $R^{-+}$. The non-spin-flip
(NSF) data, $R^{++}$ and $R^{--}$ depend on structural as well as
magnetic contribution and provide the magnetization parallel to the
applied field [($R^{++}-R^{--}) \propto <\mu_{y}>$], however, the
spin-flip (SF) intensities, $R^{+-}$ and $R^{-+}$ reflectivities
(usually, these two are degenerate) are purely of magnetic origin
and depend on the average square of the transverse in-plane magnetic
moment, $<\mu_x^2>$ \cite{zabel, sg}. One important feature of PNR
is that the value obtained for $<\mu_y>$ can be calibrated in
$\mu_B$ units because the number of scatterer can be estimated from
simultaneous analysis of R$^{++}$ and R$^{--}$ using Eq.
\eqref{pnr}. The PNR data, collected in this geometry, are
insensitive to the out-of-plane moment $\mu_z$. In PNR study of
antiferromagnetic thin-films, one obtains additional half-order
peaks corresponding to double the lattice spacing in real space of
similar spins.
 \\
\textsl{Comparison between X-ray and neutron refractivities:} The
nature of interaction for X-ray and neutron with the scatterer is
different as X-ray sees the electron cloud within the sample
(long-range electro-magnetic interaction) but neutron sees the
nucleuses as point scatterer (short-range strong interaction) in
addition with  the magnetic (long-range electro-magnetic) structure
of the scatterer (in case of polarized neutron reflectivity
utilizing the intrinsic dipole moment of neutrons). Even with
laboratory-sources of X-ray, the reflectivity as low as $10^{-8}$
and $q_z$ value as high as $\simeq$1 {\AA}$^{-1}$ can be achieved,
however for neutron reflectivity, it is difficult to have $R$ values
below $10^{-6}$ or $q_z$ beyond 0.2 {\AA}$^{-1}$. Unlike monotonic
dependence of electron density over atomic number in case of X-ray,
the neutron reflectivity provides reacher structure since the
scattering length density varies drastically from element to element
even for different isotopes of the same atom and can be negative as
well.

\textbf{Auto-correlation function and the phase problem:}\\
Here, we discuss the difficulty of finding unique ED profile from
reflectivity data. As the reflectivity expression is obtained as
modulus-squared in the final step, the information of the phase of
the scattered wave is lost in the reflectivity experiments as a
result the direct inversion of the data to reconstruct the EDP is
impossible. However, supplemented with additional information like
prior knowledge of number of layers \textit{etc.} from sample
preparation makes it possible to extract the EDP.\\

Now let us discuss the lack of uniqueness of EDP mathematically by
considering the auto-correlation function or ACF of $f(z)$, which
provides a model-independent real-space representation of the
information contained in the intensity of reflectivity pattern,
defined as,
\begin{equation}
ACF(z) = \int_{-\infty}^{\infty}f(t)^*f(z+t)dt =
\int_{-\infty}^{\infty}|F(q)|^2e^{izq}dq \label{acf}
\end{equation}
where $F(q)$ is the FT of $f(z)$. Considering
$d\rho/dz=\rho^{\prime}(z)\equiv f(z)$ and comparing eqn.
(\ref{acf}) with eqn. (\ref{Born1}), one can write, the ACF of
$\rho^{\prime}(z)$ as follows:
\begin{equation}
ACF(\rho^{\prime}(z)) = \mathcal{F}^{-1}
\left[\frac{\rho_s^2R(q_z)}{R_F(q_z)}\right] = \mathcal{F}^{-1}
\left[\mathfrak{R}(q_z) \right] \label{patterson}
\end{equation}
where $\mathcal{F}^{-1}$ is the inverse FT and
$\mathfrak{R}(q_z)=\frac{\rho_s^2R(q_z)}{R_F(q_z)}$. The above
expression is known as the Patterson function and imply the
information regarding the depth of interfaces independent of any
model. However, there is an inherent discrepancy of this data
inversion technique due to the so-called phase-loss ambiguity. As a
result some peaks may get introduced additionally or get overlooked
in the ACF of $\rho^{\prime}(z)$ as pair-wise depth difference does
matter in its calculation. A pair of sharp peaks, say, at $z_1$ and
$z_2$ with amplitudes $A_1$ and $A_2$, respectively, contributes to
a symmetric pair of very sharp peaks at $\pm(z_1-z_2)$ with
amplitude $A_1A_2$ to $ACF(z)$ along with an additive of
$A^2_1+A^2_2$ to the ACF at the origin. For $n$ interfaces
\textit{i.e.} $n$ number of peak in $\rho^{\prime}(z)$ will
pair-wise contribute to $n(n-1)$ number of peaks towards its ACF in
general. The peak for air-sample interface at $z=0$ in
$\rho^{\prime}(z)$, when considered pair-wise with other interfaces,
will contribute to $(n-1)$ peaks at their respective positions. But,
the additional peaks with the possibility of overlap and/or exact
cancelation may complicate the analysis unless one has prior
information of layers from sample growth.\\
\begin{figure*}
\begin{center}
\includegraphics[width=15 cm]{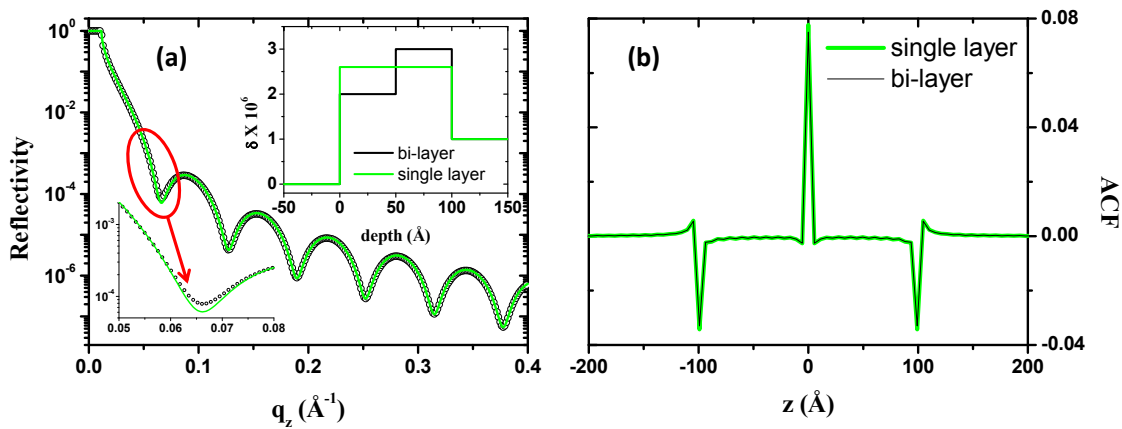}
\caption{\label{acf1} (a) A particular combination of thicknesses
and EDs may result in similar reflectivity curves when corresponding
ACFs are the same. The reflectivity curves for a bi-layer and a
single layer film of same total thickness of 100 {\AA} on the same
substrate look alike. The ED is shown in the inset and small
deviation near the first deep is magnified for clarity. (b) The
corresponding ACFs or the Patterson functions are the same in both
cases.}
\end{center}
\end{figure*}

As for example, let us consider two x-ray reflectivity curves
looking exactly the same except a small difference near the first
deep, marked by the red circle in Fig. \ref{acf1}a. The black curve
corresponds to a bi-layer film ($\delta_1 = 2\times10^{-6}$ and
$\delta_2 = 3\times10^{-6}$) having same thickness of 50 {\AA} each
on a substrate ($\delta_s = 1\times10^{-6}$). For green curve, there
is only a single layer ($\delta = 2.6\times10^{-6}$) of thickness
100 {\AA} on the same substrate. At $z=\pm 50$ {\AA}, exact
cancelation of the contributions from pairs of $\delta$-functions at
($z=0$; $z=50$ {\AA}) and ($z=50$ {\AA}; $z=100$ {\AA}) in case of
the bi-layer film results in the same ACF as for single layer, and
hence both reflectivity curves look exactly the same (refer to Fig.
\ref{acf1}b)! Interestingly, such situation may happen when
$\delta_1=\sqrt{\delta(\delta-\delta_s)}=\delta_2-\delta_s$ where
$\delta_2>\delta>\delta_1>\delta_s$.

To generate the ACF, one may directly use the discrete inverse FT
formula given by,
\begin{equation}
ACF(z)=\frac{1}{N} \sum_{k=0}^{N-1}\mathfrak{R}(q_k)e^{-iq_k k}
\end{equation}
where $k=0, 1, 2, ...., N-1$ which needs the $\mathfrak{R}(q_z)$ to
be defined from $q_{min}=0$ with uniform $\delta q_z$ and the
obtained ACF will be automatically scaled having uniform interval of
unity ($\delta z =1$). However, the truncation effect due to finite
span or window of $q_z$ which introduces ripples in the ACF with a
characteristic wavelength of $2/q_{max}$ will be unavoidable. The
peaks in ACF will be distinct and better when $q_{max}$ as large as
$\pi$ and larger the number of data points. When a standard inverse
FT tool is used, it is convenient to interpolate the reflectivity
data from $q_{min}=0$ to $q_{max}=\pi$ with uniform $\delta q=\pi/N$
so that ACF (exchanging the positive half and the negative half)
will be automatically scaled with unit interval between $\pm N$.

\textbf{Data inversion technique:} A direct data inversion is
cumbersome due to many practical difficulties like limited extent of
measured value of $q_z$, statistical noise, incoherent background
signal and the blurring from instrumental resolution.\\
Having preliminary idea about the system one can simulate the
reflectivity curve close enough to the experimental one by proper
choice of parameters. For a close choice of model EDP $\rho_m (z)$
with corresponding simulated reflectivity, $R_m (q_z)$ one may start
iteration\cite{fitmilan} using the following ansatz

\begin{equation}
\rho_e^{\prime} (z) =
\mathcal{F}^{-1}\left[\sqrt{\frac{R_e(q_z)}{R_m(q_z)}}\mathcal{F}[\rho_m^{\prime}(z)]\right]\label{fit}
\nonumber
\end{equation}

where $R_e (q_z)$ is the experimental reflectivity and
$\rho^{\prime}_e (z)$ is used as the $\rho^{\prime}_m (z)$ in the
next iteration.

\section{LabVIEW-based versatile reflectivity simulator:}
\begin{figure*}
\includegraphics[width=14 cm]{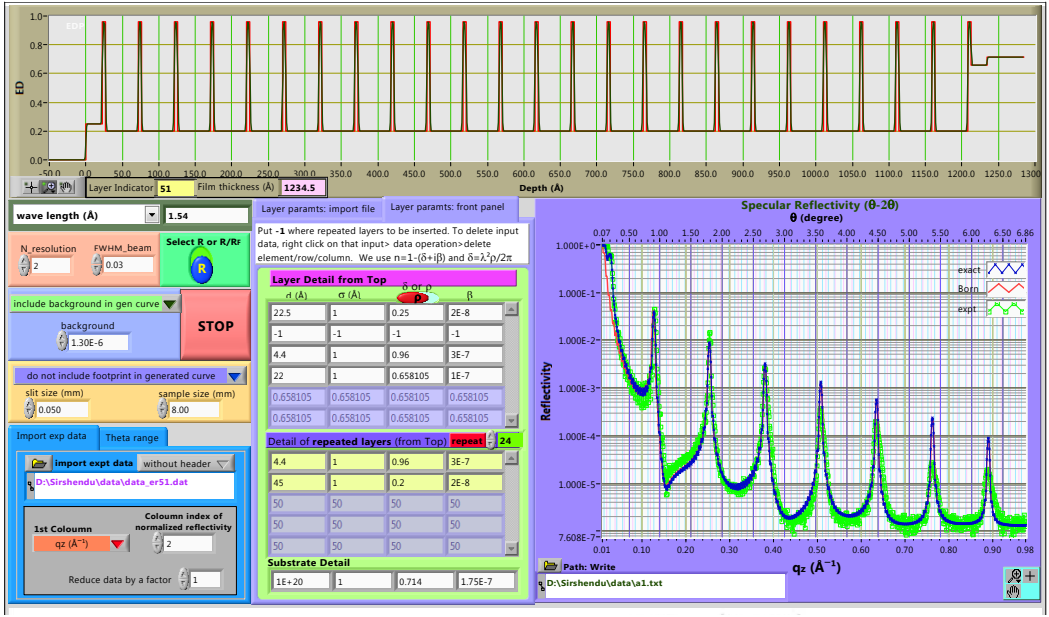}
\caption{\label{refl_labview1} Front panel of the developed specular
reflectivity simulator using LabVIEW (version $8.5$). The programme
simultaneously generates reflectivity curves for a multilayer thin
film following Born approximation (red curve) as well as the exact
dynamical theory according to the range and steps of the imported
experimental data (green) or as provided from the front panel for
model parameters (thickness, roughness, ED, absorption coefficient)
fed from the front panel or from parameter file (\textit{.txt} or
\textit{.dat} having four coloumns with headers for thickness,
roughness, delta and beta, respectively) providing the path.}
\end{figure*}

\begin{figure*}
\includegraphics[width=13 cm]{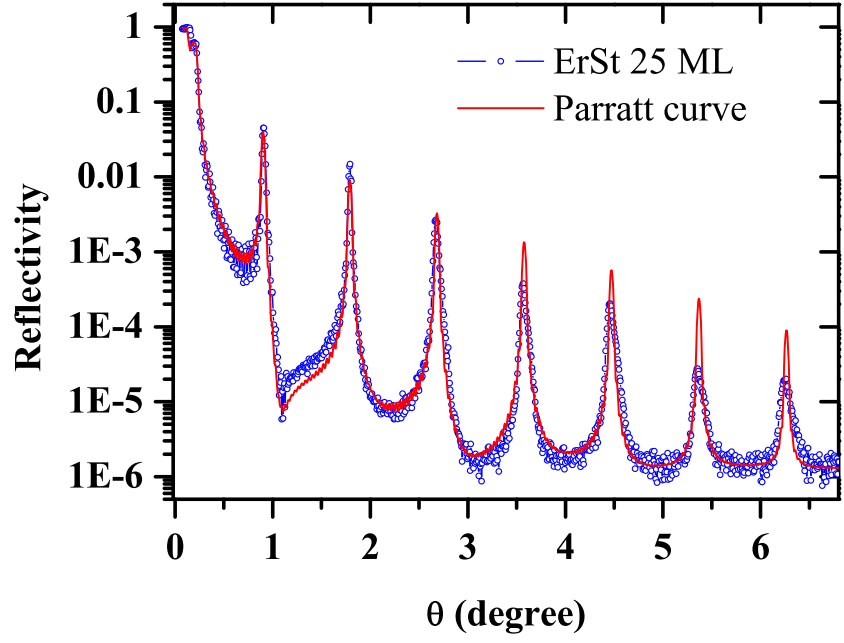}
\caption{\label{erst} The specular reflectivity curve (blue dots) of
stacked multi-layer of erbium stearate LB film on Si substrate. The
red line is the generated curve following the Parratt formalism.}
\end{figure*}

\begin{figure*}
\includegraphics[width=13 cm]{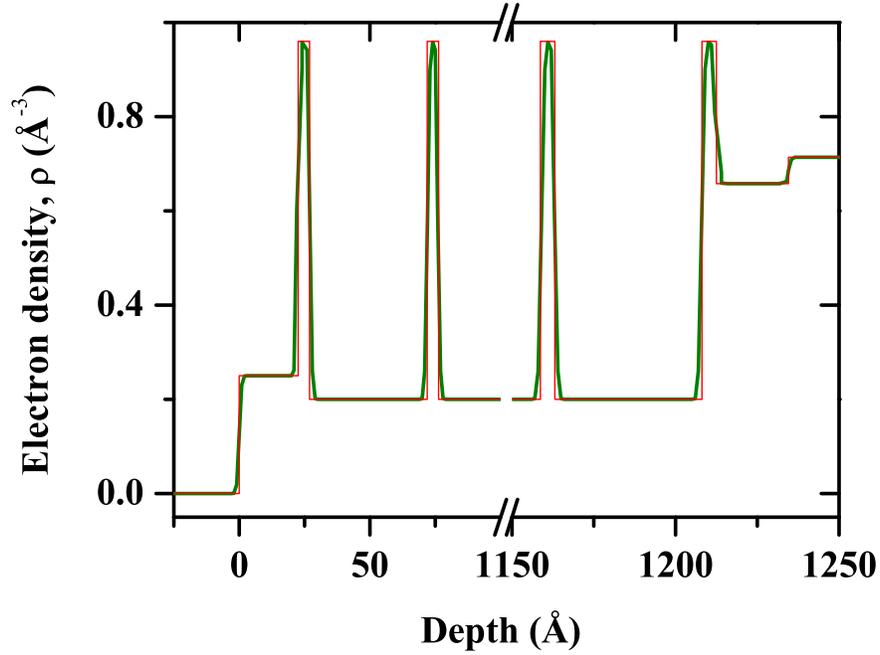}
\caption{\label{edp} Variation of electron density along the depth
[from air (z = 0) to substrate] of an erbium stearate LB film
(having 25 repetition of Er$^+$ layers, each separated by a
hydrocarbon-tail) on Si substrate with (dark green) and without
(red) roughness (Gaussian) effect.}
\end{figure*}

\begin{figure*}
\includegraphics[width=13 cm]{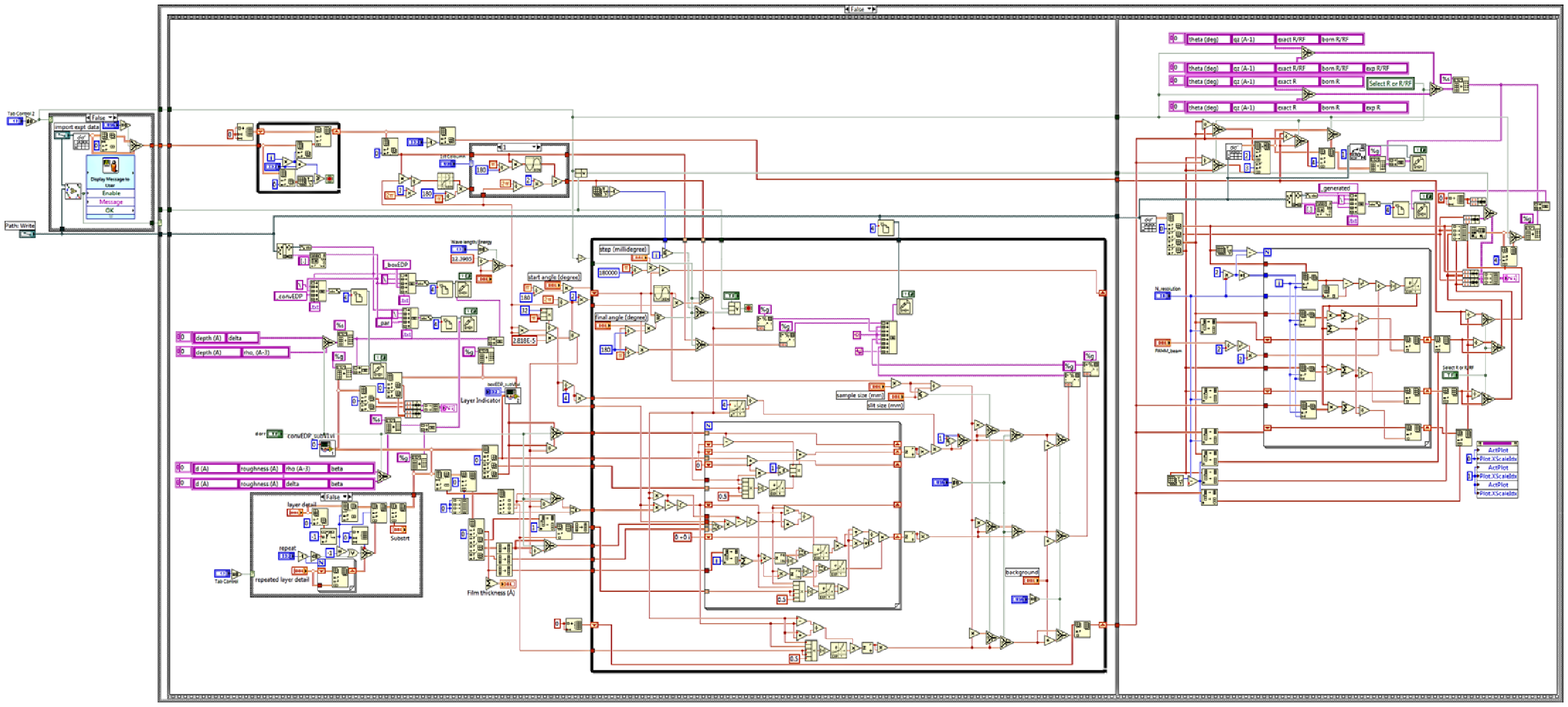}
\caption{\label{refl_labview2} Block diagram or the source code of
the developed specular reflectivity simulator using LabVIEW.
Different icons corresponding to different logical operations,
connected through wires through which data flow. }
\end{figure*}

Fig. \ref{refl_labview2} shows the block diagram of the simulator.
The code is build by drag-n-drop the icons and connecting them by
wires through which data flow following the desired logic. One can
easily modify and customize any of such programmes according to
their need and preferences.

Fig. \ref{refl_labview1} shows the screen-shot of the front panel of
the simulator which simultaneously plots specular reflectivity
curves for a multi-layer film following the (i) exact dynamical
theory (eqn. (\ref{transmatrix}) with eqn. (\ref{reflectance2})) and
the (ii) Born Theory (eqn. \eqref{Born3} with restriction $R\leq 1$)
according to the range and steps of the imported experimental data
(normalised) or as provided from the front panel corresponding to
the model parameters (thickness, roughness, ED, absorption
coefficient) fed from front panel or imported from file providing
the path. The simulator needs the followings as the input:
wavelength ($\lambda$) in {\AA} or energy in keV, range and step of
$\theta$ (or according to the experimental data when imported,
providing the path), layer detail \textit{i.e.} $d, \sigma, \delta,
\beta$ (with option to be imported, providing the path), FWHM of the
direct beam, number of data point (N$_{resolution}$) to define the
instrumental resolution window and the output path. It is easy to
include repeated layers by just putting $-1$ for $d, \sigma, \delta,
\beta$ to a particular layer where the repeated layers are intended
to be inserted. The repeated layer detail with number of repeat can
be incorporated in another window of the front panel. To delete
input data, one has to right click on that particular input $>$ data
operation $>$ delete element$/$row$/$column. Inclusion of background
and footprint effect (providing the sample size and beam width) are
optional. $R$ (or $R/R_F$ as selected) \textit{vs.} $q_z$ and $R$
(or $R/R_F$ as selected) \textit{vs.} $\theta$ along with the
electron density profile (EDP) including the roughness modification
are simultaneously plotted during run as shown in Fig.
\ref{refl_labview1}. The total thickness of the film and the number
of layers in between air and substrate is displayed in the front
panel  during run. The simulator generates five output files,
namely, the parameter file (\textit{filename}par.txt), box EDP
(\textit{filename}\_boxEDP.txt) and the convoluted EDP
(\textit{filename}\_convEDP.txt), the simulation data file with
(\textit{filename}\_generated.txt) and without
(\textit{filename}.txt) inclusion of the instrumental resolution
effect. While simulating the reflectivity curve for an experiential
data, one can reduce the simulation time using the option of
reducing the number of data points by a given factor.

 A typical reflectivity data (refer to Fig.
\ref{erst}) of a multi-layer of erbium stearate Langmuir-Blodgett
film on Si substrate obtained from a lab-source (rotating copper
anode, Enraf Nonius, model FR591) using Cu$\alpha$ characteristic
X-ray is considered here to estimate the parameters by simulating
the reflectivity curve. In the out-of-plane direction, the film
consists of 25 layers of Er separated by organic spacer (stearate
tail). The thickness of different layers from the top (\textit{i.e.}
from air to substrate), used to simulate the curve is as follows:
stearate tail (22.5 {\AA}) + 24 $\times$ [Er$^+$-head (4.4 {\AA}) +
stearate tail (45 {\AA})] + Er$^+$-head (4.4 {\AA}] + SiO$_2$ (22
{\AA}). The total film-thickness is 1234.5 {\AA} and total number of
model-layers is 51. Corresponding electron density profile with and
without the roughness effect is shown in Fig. \ref{edp}.

Once having a close guess of the parameters, experimental data can
be fitted utilizing LabVIEW platform with constrained non-linear
least-square fit option using the Levenberg-Marquardt algorithm or
the trust-region dog-leg algorithm to optimize the set of parameters
for the best fit. LabVIEW-based fitting part is in progress. A
model-independent ACF or the Patterson function generation programme
from the specular reflectivity data which provides an idea about the
thickness of layers and the depth of interfaces is also developed.
Some programmes are also developed those are useful in furnishing
the spec-files. It may be mentioned here that the stand-alone
executable version of the simulator is also built and it needs only
the LabVIEW run-time environment (free to download from National
Instruments).

\begin{acknowledgements}
I would like to acknowledge my supervisor Prof. Milan K. Sanyal for
teaching me the fundamentals of reflectivity technique, detail of
analysis and providing me the opportunity to carry out scattering
measurements.
\end{acknowledgements}

\textit{\textbf{Download link:}} One needs to run the main programme
\textit{specular reflectivity simulator.vi} only, however, it needs
two sub-programmes namely, \textit{boxEDP\_subVI.vi} and
\textit{convEDP\_subVI.vi} (to be placed in the same folder) during
run.
\href{https://docs.google.com/file/d/0B54_HAU9fQRsQ190S1h1TDRTSGM/edit?usp=sharing}{Click
here to download the zipped folder} containing these three programme
and other related programme files. One may use the stand-alone
version \textit{reflgen.exe} as well. Without having full Labview
software, one needs to install only the LabVIEW run-time engine
(free to download from National Instruments). The limitation of this
version is that the source code \textit{i.e.} the block diagram is
not available hence can not be edited. After downloading one should
run it to open the actual Labview simulator. From \textit{File} $>$
\textit{VI properties}, one can find the location of the programme
to copy and paste it at the desired folder. The programme named
\textit{BornReflGen\_ACF} can be used to generate the Born
reflectivity and corresponding ACF.  To merge reflectivity data,
furnish spec-files and for conversion of electron density, the other
programmes may be helpful.

\end{document}